\documentclass{ws-ijmpd}

\begin{document}

\def\Journal#1#2#3#4{{\it #1} {\bf #2}, (#3) #4}
\def\RPP{{Rep. Prog. Phys}}
\def\PRC{{Phys. Rev. C}}
\def\PRD{{Phys. Rev. D}}
\def\PRB{{Phys. Rev. B}}
\def\PRA{{Phys. Rev. A}}
\def\ZPA{{Z. Phys. A}}
\def\NPA{{Nucl. Phys. A}} 
\def\JPA{{J. Phys. A }}
\def\PRL{{Phys. Rev. Lett.}}
\def\PR{{Phys. Rept.}}
\def\PLB{{Phys. Lett. B}}
\def\AP{{Ann. Phys. (N.Y.)}}
\def\EPJA{{Eur. Phys. J. A}}
\def\EPJC{{Eur. Phys. J. C}}
\def\EPL{{Eur. Phys. Lett.}}
\def\NPB{{Nucl. Phys. B}}  
\def\RMP{{Rev. Mod. Phys.}}
\def\IJMPE{{Int. J. Mod. Phys. E}}
\def\AJ{{Astrophys. J.}}
\def\AJL{{Astrophys. J. Lett.}}
\def\AJSS{{Astrophys. J. Suppl. Ser.}}
\def\AA{{Astron. Astrophys.}}
\def\ARAA{{Annu. Rev. Astron. Astrophys.}}
\def\MPLA{{Mod. Phys. Lett. A}}
\def\ARNPS{{Annu. Rev. Nucl. Part. Sci.}}
\def\LRR{{Living. Rev. Rel.}}
\def\CQG{{Class. Quantum Gravity}}
\def\RAS{{Mon. Not. R. Astron. Soc.}}
\def\JCAP{{J. Cosmol. Astropart. Phys.}}
\def\GC{{Gravitation and Cosmology}}
\def\NC{{Nuovo Cimento B }}
\def\GRGC{{Gen. Relativ. Gravit.}}
\def\JMP{{J. Math. Phys.}}
\def\IJTP{{Int. J. Theor. Phys. }}
\def\JHEP{{J. High. Energy. Phys.}}
\def\IJMPD{{Int. J. Mod. Phys. D}}

\markboth{I. H. Belfaqih, H. Maulana, and A. Sulaksono}
{White dwarfs and Generalized uncertainty principle}

\catchline{}{}{}{}{}

\title{ White dwarfs and generalized uncertainty principle  }

\author{\footnotesize I. H. Belfaqih}

\address{IoT and Physics Lab, Sampoerna University, Jakarta 12780, Indonesia\\
Departemen Fisika FMIPA Universitas Indonesia, Kampus UI, Depok 16424, Indonesia
           }

\author{\footnotesize H. Maulana, A. Sulaksono}

\address{Departemen Fisika FMIPA Universitas Indonesia, Kampus UI, Depok 16424, Indonesia}

\maketitle

\begin{history}
\received{(received date)}
\revised{(revised date)}
\end{history}

\begin{abstract}
This work is motivated by the sign problem in a logarithmic parameter of black hole entropy and the existing more massive white dwarfs than the Chandrasekhar mass limit.  We examine the quadratic, linear, and linear-quadratic generalized uncertainty principle (GUP) models within the virtue of recent masses and radii of white dwarfs. We consider the modification generated by introducing the minimal length on the degenerate Fermi gas equation of state (EOS) and the hydrostatic equation. For the latter, we applied Verlinde's proposal regarding entropic gravity to derived the quantum corrected Newtonian gravity, which is responsible for modifying the hydrostatic equation. Through the models' chi-square analysis, we have found that the observation data favor the quadratic dan linear GUP models without mass limit. However, for the quadratic-linear GUP model, we can obtain the positive value of the free parameter $\gamma_0$ as well as we can get mass limit more massive than the Chandrasekhar limit. In the linear-quadratic GUP model, the formation of stable massive white dwarfs than the Chandrasekhar limit is possible only if both parameters are not equal.  
\end{abstract}

\section{Introduction}
\label{sec_intro}
Various candidates for the quantum theory of gravity predicted a minimum measurable length. Loop quantum gravity indicated that space behaves discretely at the Planck scale by defining the geometrical operator for the area, volume, and length \cite{rovelli1995discreteness,ashtekar1997quantum1,bianchi2009length,thiemann1998length,rovelli1998loop}. String theory and doubly special relativity (DSR) \cite{veneziano1986stringy,amati1989can,magueijo2002lorentz,magueijo2005string} also suggested that the Heisenberg uncertainty principle is deformed to incorporate the minimum uncertainty in the position. The deformation should provide an observational impact. However, the effect becomes insignificant at energies much below the Planck scale. This fact leads to the generalized formulation of the uncertainty principle (dubbed as GUP). See Ref.\cite{NasserTawfik} for a recent review on GUP.

The minimal measurable length from uncertainty relation assumes the Planck constant to be momentum dependent $\hbar\rightarrow\hbar(p)$. The early GUP model uses the quadratic term in momentum \cite{kempf1995hilbert,maggiore1993algebraic,brau1999minimal} as a deformation factor, which results in a minimal length and non-commutative positions relation. This model is known as the quadratic GUP model. The implications of such a modification form on the UV/IR momentum behavior, density of states, and cosmological constant problem are discussed in Ref.\cite{chang2002effect}. More recently, different GUP models have been proposed Ref.\cite{ali2009discreteness}, which contains linear and quadratic terms in momentum. The model is more general than the one in Ref. \cite{kempf1995hilbert}, and this model also ensures the fundamental commutative relation $[x_i, x_j]=[p_i, p_j]=0$. The model is used to study a single particle confined in a box of length $L$ in \cite{ali2009discreteness,ali2011proposal}, and they obtained the signature that length can be quantized. There also have been many proposals to test GUP and set their parameter bound \cite{marin2013gravitational,Villalpando1,Villalpando2,Das2021,Scardigli2019}.

Heuristically argued that GUP leads to a modified Hawking temperature of black holes \cite{adler2001generalized}. By assuming this altered temperature expression originates from the Schwarzschild metric's perturbation using the  Eddington-Robertson expansion, Scardigli-Casadio \cite{scardigli2015gravitational} obtain the upper bounds for the deformation free parameter by comparing the result with weak field observations. GUP models also yield deformed entropy-area relations. The leading correction has the logarithmic form. Various candidates of quantum gravity theories  also have predicted logarithmic and power-law correction terms to this entropy-area relation~\cite{majumder2011black,kaul2000logarithmic,medved2004conceptual,meissner2004black}. The general structure of the entropy-area relation in a black hole can be written as~\cite{Sadjadi2010}
\begin{equation}
S_{BH} = S_0+\tilde{\alpha}_0 \ln S_0-\sum_{i=1}^{\infty}\frac{ \tilde{\alpha}_i}{S_0},
\label{BHentropy}
\end{equation}
where $S_0 \approx \frac{A_h}{l_{P}^2}$ is Bekenstein-Hawking entropy of classical black holes, while the second and the higher-order terms are the logarithmic and power-law terms due to quantum corrections. The parameters $\tilde{\alpha}_i$ are finite constants. From the smallness of $l_{P}^2$, it is obvious that the dominant contribution of quantum corrections comes from the logarithmic term.

Motivated by black hole thermodynamic, the gravity and thermodynamics could be related\cite{jacobson1995thermodynamics,padmanabhan2002classical}. Verlinde \cite{verlinde2011origin} argues that gravity is an emergent phenomenon.
Contrary to the previous approaches \cite{jacobson1995thermodynamics,padmanabhan2002classical}, Verlinde's formulation is independent of the presence of horizons. By assuming that the bulk information is distributed on the surface (called holographic screens) with the entropy satisfying Bekenstein's law and that the bits of information in thermal equilibrium, he obtained Newton's law of gravity. Nicolini~\cite{nicolini2010entropic} extend Verlinde's calculation to a more general case. Several studies provide criticisms of the fundamental of Verlinde's proposal\cite{dai2017inconsistencies,Visser2011}. The corresponding authors argued that since gravity is a conservative force, then the process related to gravity itself must be reversible. Another criticism was on Verlinde's assumptions about the thermodynamic properties of the holographic screens, which in general cannot be considered as thermodynamic objects\cite{wang2018surfaces}.  However, many studies have reported that Verlinde's idea still makes sense \cite{li2010quantum,gao2010modified,kowalski2010note,chen2017entropic,cai2010notes}. For instance, Ref. \cite{kowalski2010note} shows that $SO(4, 1)$ BF theory coupled to particles can be considered the microscopic description of Verlinde's proposal. Interestingly, at Ref.~\cite {chen2017entropic} it was argued that since Verlinde's assumptions depend on the existence of a minimal length, it makes sense to use GUP in the formulation. They also obtained quantum corrections to Newton's law of gravity. On the other hand, the Planck constant dependency on momentum implies modifying the smallest possible volume in phase space. The authors of Ref. \cite{chang2002effect} obtained the modified phase space volume for the quadratic GUP model \cite{kempf1995hilbert}, while the same discussion but for the model proposed in Ref.\cite{ali2009discreteness} are reported in Ref.\cite{vagenas2019linear}. The latter will impact the equation of state (EOS) of neutron stars or white dwarfs.

Recently, the implications of GUP is on the mass-radius relation for the white dwarfs have been studied in  Refs. \cite{rashidi2016generalized,ong2018generalized}. They concluded that by including GUP on the degenerate gas EOS, the mass and radius increase unboundedly for the deformation parameter's positive sign, and therefore, the Chandrasekhar limit ceases to exist \cite{ong2018generalized}. For the GUP parameter's negative value, the mass is bound to Chandrasekhar mass as the radius increases unboundedly. These findings can be understood since GUP forbids an object from collapsing to zero sizes. Authors of Ref. \cite{mathew2018effect} examine GUP's effect by analytically solving the modified white dwarf EOS due to GUP using the Lane-Emden equation to obtain the mass-radius relation. They found that by increasing the central Fermi momentum $p_{c}$, the radius approaches a minimum value.
In contrast, the mass increases slowly, but after reaching a particular $p_c$ value, the mass and radius increase boundlessly. This behavior is due to the saturated value of particle density when $p_c\rightarrow\infty$. From this fact, one can conclude that at very high Fermi momentum, the white dwarf behaving like a degenerate star with a constant density. Recently, Ref.\cite{mathew2020existence} using the Tolman-Oppenheimer-Volkoff (TOV) equation instead of Newtonian hydrostatic to discuss white dwarfs mass-radius in the GUP framework. They obtained Chandrasekhar's mass limit and a stable mass-radius relation for the positive value GUP parameter. From their stability analysis of the corresponding gravitational collapse beyond  Chandrasekhar's mass limit, the formation of compact objects denser than white dwarf is possible.  However, we need to highlight here that the previous works \cite{rashidi2016generalized,ong2018generalized,mathew2018effect,mathew2020existence} do not take into account the impacts of the presence of minimal length that could also generate a deformation in the Newtonian potential. Therefore, it seems that the actual cause of the existence of mass limit within GUP models is still needed to be scrutiny.

Related to white dwarf properties, we need to note some recent observations and theoretical progress. The type Ia supernova (SNe Ia) has an essential role in astrophysics as standard candles to measure far distances in-universe. SNe Ia event should cause by the ultraviolet thermonuclear explosion of carbon-oxygen white dwarf when the white dwarf mass approach the Chandrasekhar mass limit~\cite{DM2015,JW2016}. Some of these SNe Ia are observed as highly over-luminous objects, e.g., SN2003fg, SN2006gz, SN2007if, and SN2009dc~\cite{Howell2006,Scalzo2010}. It implies that their progenitor could be white dwarfs with super-Chandrasekhar masses limit in the range of 2.1 -2.8 M$_\odot$~\cite{Howell2006,Scalzo2010,Hicken2007,Yamanaka2009,Silverman2011,Taubenberger2011} where M$_\odot$ is solar mass. Such a large mass limit can not be explained by using the standard theory of white dwarfs. Up to now, there are some solutions or exotic mechanisms proposed in attempting to understand these ultra massive white dwarfs (Please see Refs.~\cite{DM2015,JW2016} and the references therein.). Therefore, it is pretty interesting to check whether GUP models can also explain this exotic mass limit and also compatible with some masses and radii data of standard white dwarfs\cite{holberg2012observational}, respectively.

This work investigates the impacts of the quadratic, linear, and linear-quadratic GUP models on white dwarf properties, particularly the mass-radius relation. We evaluate twofold corrections generate by introducing GUP. First, we review modification by GUP on the EOS of the degenerate Fermi gas. Second, we explore the quantum corrections on the hydrostatic equation due to the presence of minimal length. To obtain it, we apply the entropic gravity argument to calculate the modified Newton's law of gravity, which we will use to obtain the quantum modified hydrostatic equation. The presence of GUP generates modification on the Bekenstein-Hawking formula, which is one of the crucial aspects of entropic gravity. To examine the sign of GUP parameter favor by white dwarfs properties and the compatibility GUP models with white dwarf masses and radii data from observations, we parametrize the free parameter of the corresponding GUP model to recent mass-radius of white dwarfs data\cite{holberg2012observational}. We also use a $\chi^2$ analysis to obtain each GUP model's parameter acceptable value with its confidence.

This paper is structured as follows. Sec. \ref{s2} briefly reviews some aspects of the GUP models, including the expression for the phase space volume deformation predicted by the model. This section also discusses the deformation of the Newtonian gravity by following Verlinde's method with the minimal length for the quadratic, linear, and linear-quadratic  GUP models. We provide the expression for the quantum corrected hydrostatic equation due to the corresponding non-relativistic gravitation potential deformation. Sec. \ref{s6} analyses the EOS deformation of degenerate Fermi gas due to the  GUP corrections. This section also discusses the corresponding modification on the mass-radius curve of white dwarfs based on the EoS and hydrostatic equation's deformation. We also set the bound for the GUP parameters by comparing them with the recent mass-radius data of white dwarfs \cite{holberg2012observational}. Finally, we conclude our work in Sec.\ref{s8}. Here we adopt SI units so that the calculation results will be expressed in terms of the Boltzmann constant $k_\text{B}$, the speed of light $c$, the gravitational constant $ G $, and the reduced Planck constant $\hbar$. 

\section{Theoretical Framework}\label{s2}
\label{sec_MC}
 \subsection{Models of GUP}
In this subsection we discuss briefly three forms of the momentum dependencies of the $\hbar(p)$ known in literature i.e., the quadratic \cite{kempf1995hilbert,maggiore1993algebraic,brau1999minimal}, linear \cite{ali2011minimal} and linear-quadratic \cite{ali2009discreteness} GUP models. The linear model can be considered as a first-order correction of the linear-quadratic model. Rather than taking the linear model parameter to be the square root of the parameter of the quadratic model, we assume that both parameters are independent of each other. 

For quadratic GUP in three dimensions, the commutator has the form \cite{kempf1995hilbert}
\begin{equation}
\left[x_i,p_j\right] = i\delta_{ij}\hbar\left(1+\gamma p^2\right),
\end{equation}
where $p^2$ is the magnitude of momentum 3-vector, and $\gamma$ is the deformation factor. Based on this modified commutator the uncertainty relation between position and momentum are deformed such as
\begin{equation}\label{P1}
\Delta x\Delta p\geq\frac{\hbar}{2}\left[1+\gamma\left(\Delta p\right)^2\right].
\end{equation}
By solving those inequality, one would get
\begin{equation}
\Delta x\geq \hbar \sqrt{\gamma},
\end{equation}
there is a non-vanishing minimal uncertainty in the position. The number of states in phase space based on this model is modified \cite{chang2002effect} to
\begin{equation}
\int d^3xd^3p\rightarrow\int \frac{d^3xd^3p}{\left(1+\gamma p^2\right)^3},
\end{equation}
so states with high momentum becomes fewer compare to that of the low momentum.  Another approach are considering the linear term in addition to the quadratic term \cite{ali2009discreteness,ali2011proposal}. In this model the commutator has the form 
\begin{equation}
\left[x_i,p_j\right]= i\hbar\left[\delta_{ij}-\beta\left(p\delta_{ij}+\frac{p_ip_j}{p}\right)+\beta^2\left(p^2\delta_{ij}+3p_ip_j\right)\right].
\end{equation}
Note, however, for phenomenological purpose, we take a less tight assumption in this work than that of Ref. \cite{ali2009discreteness,ali2011proposal}, where we assume that the quadratic parameter independent of the linear parameter $\gamma\neq\beta^2$ where the $x, p$ commutation relation is expressed as
\begin{equation}\label{tig1}
\left[x_i,p_j\right]\equiv i\hbar\left[\delta_{ij}-\beta\left(p\delta_{ij}+\frac{p_ip_j}{p}\right)+\gamma\left(p^2\delta_{ij}+3p_ip_j\right)\right].
\end{equation}
In the Subsec. \ref{KKl}, we will applied this form of commutator to evaluate the mass-radius relation of white dwarfs, and by fitting with the data \cite{holberg2012observational} we obtain a relation between $\gamma$ and $\beta$. This modified relation also introduce lower bound for the uncertainty in position and and also an upper bound for the momentum uncertainty
\begin{equation}
\Delta x\geq \left(\Delta x\right)_{\text{min}}\approx \hbar\left(2\sqrt{\gamma}-\beta\right),
\end{equation}
 and
\begin{equation}
  \Delta p\leq \left(\Delta p\right)_{\text{max}}\approx\frac{1}{2\sqrt{\gamma}},
  \label{aba}
\end{equation}
 with a lower bound for the position, uncertainty can be claimed as a minimum measurable distance, while upper bound of momentum uncertainty can claim that momentum measurements cannot be arbitrarily imprecise \cite{majumder2011black}. The number of states in the six-dimensional phase space within this model is also deformed to
\begin{equation}
\int d^3xd^3p\rightarrow \int \frac{d^3xd^3p}{\left[1-\beta p+\gamma p^2\right]^4}.
\end{equation}
The power that appears in the denominator differs from the quadratic GUP model. The calculation in \cite{vagenas2019linear} shows that in $D$ dimensional configuration space, the power is $D+1$ for linear quadratic deformation, while for quadratic model the power is $D$ \cite{chang2002effect}.

The discussion for the first-order correction of Eq. (\ref{tig1}), is discussed in detail on Ref.\cite{ali2011minimal}. By taking only the linear term of Eq. (\ref{tig1}):
\begin{equation}
\left[x_i,p_j\right] = i\hbar\left[\delta_{ij}-\beta\left(p\delta_{ij}+\frac{p_ip_j}{p}\right)\right],
\end{equation}
the modified number of states reduced to \cite{ali2011minimal}
\begin{equation}
  \int d^3xd^3p\rightarrow \int \frac{d^3xd^3p}{\left[1-\beta p\right]^4}.
  \label{psL}
\end{equation}
It is clear that the phase space volume in right handed in Eq. (\ref{psL}) is similar to that of the linear-quadratic GUP model, in which both models has the same power in the denominator factor of the phase space integral. This result is not surprising since the linear GUP model is just the first-order correction of the linear-quadratic GUP model.

\subsection{Modified Newtonian hydrostatic equation}
\label{s3}
In this subsection, we followed Verlinde's proposal of gravity as an entropic force \cite{verlinde2011origin} to derived the modified version of Newtonian gravity due to the GUP effect. The proposal relied on the Bekenstein's formula \cite{bekenstein1973black} relating the entropy and the surface area of the black holes. The analysis shows that black holes entropy is proportional to its surface area, while various quantum gravity models predicted correction terms \cite{kaul2000logarithmic,carlip2000logarithmic}.

From the knowledge of the black hole thermodynamics, the increase in the  area of a black hole when absorbing a classical particle with energy $E$ and size $\Delta x$ is expressed as
\begin{equation}\label{P4}
\Delta A\geq \frac{8\pi l^{2}_{P}}{\hbar c}E\Delta x\geq\frac{8\pi l^{2}_{P}}{\hbar}\Delta p\Delta x.
\end{equation}
where the Planck length is $l_P$ $\approx 1.616\times10^{-35}$m. The effect of GUP is entering through $\Delta x\Delta p$ term. The quadratic GUP modified the entropy-area relation, through Eq. (\ref{P4}). By solving Eq. (\ref{P1}) to first order in $\gamma$ and substitute the result into Eq. (\ref{P4}) one obtain 
\begin{equation}
\Delta A\geq 4\pi l^{2}_{P}\left[1+\frac{\hbar^2\gamma}{4\left(\Delta x\right)}\right].
\end{equation}
The estimation of $\Delta x$ can be written as \cite{adler2001generalized,scardigli2015gravitational}
\begin{equation}
\Delta x\simeq 2\pi R_{\text{S}},
\end{equation}
where $R_{\text{S}}$ is the unmodified Schwarzschild radius $R_{\text{S}}=\frac{2GM}{c^2}$. By using this estimation, we obtain the lower bound for increasing area to first order in $\gamma$ as
\begin{equation}
\Delta A_{\text{min}}\simeq 4\pi l^{2}_{P}\lambda\left[1+\frac{\hbar^2\gamma}{4\pi A}\right],
\end{equation}
with $A=4\pi R_{\text{S}}^{2}$ is the area of a black hole horizon, and $\lambda$ is proportional constant will be fixed later. For the entropy, the minimum increase is given by one bit, and generally, it is considered that $\Delta S_{\text{min}}=b=\ln 2$. Then according to these results, we can write
\begin{eqnarray}
\frac{dS}{dA}\simeq \frac{\Delta S_{\text{min}}}{\Delta A_{\text{min}}}&=&\frac{b}{4\lambda\pi l^{2}_{P}\left[1+\frac{\hbar^2\gamma}{4\pi A}\right]}\nonumber \\
&\approx&\frac{b}{4\pi\lambda l^{2}_{P}}\left(1-\frac{\hbar^2\gamma}{4\pi A}\right).
\end{eqnarray}
After doing integration, we obtain
\begin{equation}
S\left(A, \gamma\right)=\frac{b}{4\lambda\pi l_{P}^2}\left[A-\frac{\hbar^2\gamma}{4\pi}\ln\left(\frac{A}{4l_{P}^{2}}\right)\right].
\end{equation}
Now by demanding that when $\gamma\rightarrow0$ the relation above is reduced to the standard Bekenstein-Hawking entropy-area equation $S\left(A, 0\right)=S_{\text{BH}}$, we can identify that $\frac{b}{4\pi\lambda l_{P}^{2}}=\frac{k_{\text{B}}c^3}{4G\hbar}$. The modified entropy-area equation now can be written as
\begin{equation}\label{BH1}
S\left(A, \gamma\right)=\frac{k_{\text{B}}c^3}{4G\hbar}\left[A-\frac{\hbar^2\gamma}{4\pi}\ln\left(\frac{A}{4l^2_{P}}\right)\right].
\end{equation}
After obtaining the modified entropy-area relation, we consider Verlinde's proposal to derived the modified Newtonian gravity. The proposal relies on the holographic principle \cite{hooft1993dimensional,susskind1995world} which states, that the microscopic information (bits) of a closed surface is embedded on the surface of radius $r$ (where generally assumed the surface to be a sphere for simplicity). The number of bits can be written as
\begin{equation}\label{B}
N=\frac{4S}{k_{\text{B}}},
\end{equation}
where $k_{\text{B}}$ is the Boltzmann constant. It is conjectured in Ref. \cite{verlinde2011origin} that these bits carried the energy of the system and that the energy is divided evenly over the bits $N$. The temperature is then determined by the equipartition theorem
\begin{equation}\label{PT}
T=\frac{2E}{Nk_{\text{B}}},
\end{equation}
where $E=Mc^2$ is the total energy of the system. Motivated by Bekenstein's argument, when a test particle $m$ is approaching the surface by an amount $\Delta x$, the change of the entropy associated with the information on the boundary can be written as
\begin{equation}
\Delta S=2\pi k_{\text{B}}\frac{\Delta x}{\hbar}mc.
\end{equation}
By relating $\Delta S$  with the entropic force, then
\begin{eqnarray}\label{B2}
F\Delta x&=&T\Delta S \nonumber \\
F&=&\frac{2Mc^2}{N}2\pi \frac{mc}{\hbar} \nonumber \\
&=&\frac{4\pi Mmc^3}{\hbar}N^{-1},
\end{eqnarray}
where we have used the expression in Eq. (\ref{PT}) to replace $T$. The number of bits is defined by Eq. (\ref{B}). By substituting Eq. (\ref{BH1}) into Eq. (\ref{B}), we obtain
\begin{eqnarray}
N&=&\frac{c^3}{G\hbar}A\left[1-\frac{\hbar^2\gamma}{4\pi A}\ln \left(\frac{A}{4l_{P}^{2}}\right)\right].
\end{eqnarray}
If we substitute this result into Eq. (\ref{B2}) and then taking only the first order correction we obtain
\begin{eqnarray}\label{F1}
F_{\text{Q}}=G\frac{Mm}{r^2}\left[1+\frac{\hbar^2\gamma}{16\pi^2r^2}\ln\left(\frac{\pi r^2}{l_{P}^{2}}\right)\right].
\end{eqnarray}
To this end, we have obtained the correction term for the Newtonian gravity due to the quadratic GUP model. Therefore, we can write the gravitational field generated by the mass $M$ as
\begin{equation}\label{QQ}
\vec{g}_{\text{Q}}(r)=-\frac{GM}{r^2}\left[1+\frac{\gamma_{0}l_{P}^{2}}{16\pi^2r^2}\ln\left(\frac{\pi r^2}{l_{P}^{2}}\right)\right]\hat{r}.
\end{equation}
Here we introduce $\gamma=\frac{\gamma_{0}l_{P}^{2}}{\hbar^2}$, with $\gamma_{0}$ is free dimensionless parameter.

By doing exactly the same procedure as before, we also can obtain the deformed Bekenstein-Hawking formula for the linear-quadratic GUP model as
\begin{eqnarray}\label{BH2}
S\left(A,\beta_0,\gamma_0\right)=&&k_{\text{B}}\left(\frac{A}{4l_{P}^{2}}\right)+\frac{k_{\text{B}}\beta_0}{\sqrt{\pi}}\sqrt{\frac{A}{4l_{P}^{2}}}\nonumber \\&&-\frac{k_{\text{B}}\gamma_0}{4\pi}\ln\left(\frac{A}{4l_{P}^{2}}\right). \nonumber \\
\end{eqnarray}
Then we also can obtain the gravitational force between masses as of the linear-quadratic GUP model as
\begin{eqnarray}\label{F2}
F_{\text{LQ}}(r)=-G\frac{mM}{r^2}\left[1-\frac{\beta_{0}l_{P}}{\pi r}+\frac{l_{P}^{2}}{\pi^2r^2}\left(\beta_{0}^{2}+\frac{\gamma_{0}}{4}\ln\left(\frac{\pi r^2}{l_{P}^{2}}\right)\right)\right],\nonumber \\
\end{eqnarray}
 while the gravitational field generates by mass $M$ in the linear-quadratic model becomes
\begin{equation}\label{BiB}
\vec{g}_{\text{LQ}}(r)=-G\frac{M}{r^2}\left[1-\frac{\beta_{0}l_{P}}{\pi r}+\frac{l_{P}^{2}}{\pi^2r^2}\left(\beta_{0}^{2}+\frac{\gamma_{0}}{4}\ln\left(\frac{\pi r^2}{l_{P}^{2}}\right)\right)\right]\hat{r}.
\end{equation}
For the linear model, we can take only the $\beta_0$ correction term of Eq. (\ref{BH2}) as
\begin{eqnarray}\label{BH2W}
S\left(A,\beta_0\right)=k_{\text{B}}\left(\frac{A}{4l_{P}^{2}}\right)+\frac{k_{\text{B}}\beta_0}{\sqrt{\pi}}\sqrt{\frac{A}{4l_{P}^{2}}},
\end{eqnarray}
and we obtain the modified Newtonian gravity as
\begin{equation}\label{L33}
F_{\text{L}}(r)=-G\frac{mM}{r^2}\left[1-\frac{\beta_{0}l_{P}}{\pi r}\right].
\end{equation}
and so, the gravitational field becomes
\begin{equation}\label{L}
\vec{g}_{\text{L}}(r)=-G\frac{M}{r^2}\left[1-\frac{\beta_{0}l_{P}}{\pi r}\right]\hat{r}.
\end{equation}
The quadratic GUP model correction consists only of the attractive part, and the linear model consists only of a repulsive part, while the linear-quadratic model consists of both corrections. We see that the correction terms in GUP models is proportional to a tiny number i.e., $l_{P}^{2}\approx10^{-70}$ . In general we assume that this correction term should be neglected for large distance $r\gg l_P$, because the corrections is only important near the Planck scale i.e., $r\rightarrow l_P$. Here, we will explicitly examine this asumption.

Note that the corrections have a quantum nature through the presence of $l_{P}$, which can be written as $l^{2}_{P}=G\hbar/c^3$. An exciting remark has been made in Ref. \cite{chen2017entropic}, where the author argues that since Verlinde's proposal incorporates minimal length, then naturally, GUP should be included in the formulation. Based on this argument, they show that $\hbar$ appears as the next leading term in Newtonian gravity, as shown in our result. Interestingly two new assumptions are proposed in Ref. \cite{chen2017entropic}. First, the temperature is defined through the degrees of freedom associated with the photon that originates from the source's rest mass energy $Mc^2$. Finally, the entropy variation is followed the metric formulation of Fursaev \cite{Fursaev1,Fursaev2} applied to the surface area of a sphere. In contrast, our results here are followed directly from Verlinde's assumptions. However, only the linear-quadratic result in Eq. (\ref{F2}) have some resemblance form with Eq. (22) in Ref. \cite{chen2017entropic}, that is both contain $r^{-4}$ and logarithmic corrections to the quantum corrections of the Newtonian gravity. The appearance factor  $r^{-4}$ in our derivation is due to the presence of  $\sqrt{A}$ term as a next leading order in the Bekenstein-Hawking formula for the linear-quadratic model, while in Ref. \cite{chen2017entropic} is originated from taking into account the deformation of the defining temperature. Nevertheless, our corrections also contain factor $r^{-3}$, which came from the linear part of the modified Bekenstein-Hawking relation.

The hydrostatic equation governing the gradient of the pressure to the radial distance of the non-relativistic stars which are derived from the Newtonian-like gravity equation. Consider a star with the profile mass $M(r)$. The unmodified gravitational force $dF$ acted on the spherical shell of thickness $dr$ and surface area $A$ due to the material inside it is given by
\begin{eqnarray}\label{F}
dF=AdP=-G\frac{M(r)dM(r)}{r^2},
\end{eqnarray}
$dM(r)$ is the mass of the spherical shell, while $dP$ is the outward pressure balancing the gravitational attraction. By writing $dM(r)=\rho(r)Adr$, we can express
\begin{equation*}
  \frac{1}{\rho(r)}\frac{dP}{dr}=-\frac{GM}{r^2},
  \label{dpdr1}
\end{equation*}
where $\rho(r)$ is the mass density as a function of radius. Now by differentiating Eq. (\ref{F}) to $r$, we obtain
\begin{equation*}
\frac{d}{dr}\left(\frac{r^2}{G\rho(r)}\frac{dP}{dr}\right)=-4\pi r^2\rho(r),
\end{equation*}
here we used the mass formula  
\begin{equation*}
  \frac{dM}{dr} \equiv 4\pi r^2\rho(r).
  \label{dmdr}
\end{equation*}
It is conventional to write Eq. (\ref{F}) in term of energy density, so in this respect, we have
\begin{equation}\label{Mass1}
\frac{dM}{dr}=4\pi r^2\frac{\epsilon(r)}{c^2},
\end{equation}
while the hydrostatic equation becomes
\begin{equation}\label{hydro1}
\frac{c^4}{G}\frac{d}{dr}\left(\frac{r^2}{\epsilon(r)}\frac{dP}{dr}\right)=-4\pi r^2\epsilon(r).
\end{equation}
Eq. (\ref{Mass1}) and Eq. (\ref{hydro1}) simultaneously describe the structures of the standard non-relativistic stars such as main-sequence stars, brown dwarfs, and white dwarfs. The inclusion of GUP into the Bekenstein-Hawking equation and Verlinde's proposal modify the Newtonian gravitational field, and hence modified the hydrostatic equation. For the quadratic GUP model, the modified gravitational force is given by Eq. (\ref{F1}). By using these results, Eq. (\ref{F}) can be modified into
\begin{eqnarray}
dF=AdP=-G\frac{M(r)dM(r)}{r^2}\left(1+\frac{\hbar^2\gamma}{16\pi^2 r^2}\ln\left(\frac{\pi r^2}{l_{P}^{2}}\right)\right).\nonumber \\
\end{eqnarray}
As before by writing $dM(r)=\rho(r)Adr$, we obtain
\begin{eqnarray*}
\frac{dP}{dr}&=&-G\rho(r)\frac{M(r)}{r^2}\left(1+\frac{\hbar^2\gamma}{16\pi^2 r^2}\ln\left(\frac{\pi r^2}{l_{P}^{2}}\right)\right)\nonumber \\
\end{eqnarray*}
By re-expressing and differentiating the relation above, we obtain
\begin{eqnarray}
  \frac{16\pi^2c^4}{G}\frac{d}{dr}\left[\frac{r^4}{\epsilon(r)\left[16\pi^2r^2+\hbar^2\gamma\ln\left(\frac{\pi r^2}{l_{P}^{2}}\right)\right]}\frac{dP}{dr}\right]=-4\pi r^2\epsilon(r).\nonumber \\
  \label{MHSE}
\end{eqnarray}
Eq. (\ref{MHSE}) is the modified hydrostatic equation due to the correction from the quadratic GUP model. It becomes Ref. (\ref{hydro1}) if $\gamma$ is set to be zero. Using the same procedure, we also can obtain the modified hydrostatic equation due to the correction from the linear-quadratic GUP model as

\begin{eqnarray}
\frac{\pi^2c^4}{G}\frac{d}{dr}\left[\frac{r^4}{\epsilon(r)\left[\pi^2r^2-\hbar\beta\pi r+\hbar^2\left(\beta^{2}+\frac{\gamma}{4}\ln\left(\frac{\pi r^2}{l_{P}^{2}}\right)\right)\right]}\frac{dP}{dr}\right]=-4\pi r^2\epsilon(r),
\end{eqnarray}

and for the linear GUP model as
\begin{eqnarray}
\frac{\pi^2c^4}{G}\frac{d}{dr}\left[\frac{r^4}{\epsilon(r)\left[\pi^2r^2-\hbar\beta\pi r\right]}\frac{dP}{dr}\right]=-4\pi r^2\epsilon(r).
\end{eqnarray}
These modified hydrostatic analytic results are needed to obtain the mass-radius relation for the white dwarfs. 
\section{Results and Discussion}\label{s6}
For the energy near the Planck scale, GUP models predict the corrections for compact objects' thermodynamics properties. According to this model, this correction comes from that near the Planck scale, the Planck constant is not constant, but a function of momentum $\hbar\left(p\right)$. Since the usual quanta volume of the phase space is $\hbar^3$, in the presence of GUP, the quanta volume must be transformed into $\hbar(p)^3$. The authors of Ref. \cite{chang2002effect} have found the $\hbar(p)^3$ term correction within the quadratic GUP model. While the authors of Refs.  \cite{vagenas2019linear,ali2011minimal} have found the power of the momentum in the correction factor is four instead of three in the linear-quadratic and linear GUP models. This section discusses how the GUP models modified the expression for the several thermodynamics quantities in the zero-temperature ground state of white dwarf EOSs due to the change in the phase space volume definition. First, we discuss EOS analytically with stressing in the large limit of Fermi momentum, and after that, we discuss the compatibility of GUP models with recent observation data of white dwarfs masses and radii. 

\subsection{Degenerate Fermi gas}\label{jiah}
 The number of states for the quadratic GUP has the form
\begin{equation}
\Omega=\frac{g}{\left(2\pi\right)^3\hbar^3}\int \frac{d^3xd^3p}{\left(1+\gamma p^2\right)^3},
\end{equation}
where $g$ is the spin degeneracy. As already discussed in Ref.\cite{mathew2018effect}, this phase space volume modified the  number density, pressure, and energy density for the degenerate Fermi gas as
\begin{eqnarray}
n\left(p_F\right)&=&\frac{g}{2\pi^2\hbar^3}\int_{0}^{p_F}\frac{p^2dp}{\left(1+\gamma p^2\right)^3}, \\
P\left(p_F\right)&=&\frac{g}{2\pi^2\hbar^3}\int_{0}^{p_F}\frac{p^2dp}{\left(1+\gamma p^2\right)},\nonumber \\&&\times\left[\sqrt{p_{F}^2c^2+m^2c^4}-\sqrt{p^2c^2+m^2c^4}\right]\\
\epsilon\left(p_{F}\right)&=&\frac{g}{2\pi^2\hbar^3}\int_{0}^{p_F}\frac{p^2\sqrt{p^2c^2+m^2c^4}dp}{\left(1+\gamma p^2\right)^3},
\end{eqnarray}
where $p_F$ is the Fermi momentum, this modification provides a different asymptotic behavior of the corresponding EOS. In the standard model of Fermi gas, the number density approaches infinity when the Fermi momentum goes to infinity since $n\propto p_{F}^3$. The quadratic GUP model, on the other hand, gives a particular saturation value for $n$ when $p_{F}\rightarrow\infty$ to
\begin{eqnarray}
n\left(\infty\right)=\frac{\pi^2}{2\left(2\pi\right)^3\hbar^3\gamma^{\frac{3}{2}}}.
\end{eqnarray}
This value is because the GUP model forbids the Fermi gas from collapsing to zero sizes. At this large limit of Fermi momentum ($p_F\rightarrow\infty$), the average distance between Fermi particles is constant such as
\begin{eqnarray}
\bar{d}=n^{-\frac{1}{3}}\propto\hbar\beta\sim l_{P}.
\end{eqnarray}
The exact value of $\bar{d}$ is around Planck length. Note that the pressure increases more slowly in the quadratic GUP model with the increasing Fermi momentum. While the standard ideal Fermi gas gives $P\propto p_{F}^{4}$ for very large values of $p_{F}$, and the quadratic GUP model predict $P\propto p_{F}$. This effect is expected not too significant in the typical range of the number density of ideal Fermi gas EOS in white dwarfs. However, we hope that a small physical signature remnant of GUP can still be imprinted in white dwarf properties that might explain the existence of white dwarfs with a mass larger than the Chandrasekhar mass limit. For completeness in the following, we also discuss the thermodynamical properties of linear-quadratic and linear GUP models.  

For the linear-quadratic GUP model, the number of states has the form \cite{vagenas2019linear,ali2011minimal}
\begin{equation}\label{lqg}
\Omega=\frac{g}{\left(2\pi\right)^3\hbar^3}\int \frac{d^3xd^3p}{\left(1-\beta p+\gamma p^2\right)^4}.
\end{equation}
Here we described the behavior of the degenerate Fermi gas under this GUP model.The number of particles $N$ is given by
\begin{eqnarray*}
N=2\sum_{p=0}^{p_F}N_p=2\frac{1}{\left(2\pi\right)^3\hbar^3}\int\int_{0}^{p_{F}}\frac{d^3xd^3p}{\left(1-\beta p+\gamma p^2\right)^4},
\end{eqnarray*}
which directly yields the number density $n=\frac{N}{V}$ as

\begin{eqnarray}\label{den}
n\left(x_F\right)&=&\frac{1}{\pi^2\hbar^3}\int_{0}^{p_{F}}\frac{p^2dp}{\left(1-\beta p+\beta^2p^2\right)^4}\nonumber \\
&=&\frac{1}{\pi^2\hbar^3}\left[-\frac{\beta^3+26\beta\gamma}{3\left(\beta^2-4\gamma\right)^3}+\frac{8\gamma\left(\beta^2+\gamma\right)\tan^{-1}\left(\frac{\beta}{\sqrt{-\beta^2+4\gamma}}\right)}{\left(-\beta^2+4\gamma\right)^{\frac{7}{2}}}-\frac{1}{3\left(\beta^2-4\gamma\right)^3}F\left(\beta, \gamma, p_{F}\right)\right],\nonumber \\
\end{eqnarray}

with $F\left(\beta, \gamma, p_{F}\right)$ defined as
\begin{eqnarray}
F\left(\beta, \gamma, p_{F}\right)&=&-\frac{6\left(\beta^2+\gamma\right)\left(\beta-2\gamma p_{F}\right)}{\left(1-\beta p_{F}+\gamma p_{F}^{2}\right)}+\frac{\left(\beta^2-4\gamma\right)\left(\beta^2+\gamma\right)\left(\beta-2\gamma p_{F}\right)}{\gamma\left(1-\beta p_{F}+\gamma p_{F}^{2}\right)^2}
\nonumber \\
&&+\frac{\left(\beta^2-4\gamma\right)^2\left(-\beta+\beta^2p_{F}-2\gamma p_{F}\right)}{\gamma\left(1-\beta p_{F}+\gamma p_{F}^{2}\right)^3}+\frac{24\gamma\left(\beta^2+\gamma\right)\tan^{-1}\left(\frac{-\beta+2\gamma p_{F}}{\sqrt{-\beta^2+4\gamma}}\right)}{\sqrt{-\beta^2+4\gamma}}.\nonumber\\
\end{eqnarray}
For the standard Fermi gas EOS, the relation between number density and the Fermi momentum is just a simple cubic function, in which $n\propto p_{F}^{3}$. Due to the modification of the phase space volume, the relation between them should be expressed by a transcendental function.  Therefore, we cannot express the Fermi momentum analytically in terms of number density. The expression for the degenerate energy can be written as
\begin{eqnarray}
E=2\sum_{p=0}^{p_{F}}\epsilon_{p}&=&\int\int_{0}^{p_F}\frac{d^3xd^3p\epsilon_{p}}{\left(2\pi\right)^3\hbar^3\left[1-\beta p+\gamma p^2\right]^4},\nonumber \\
\end{eqnarray}
which eventually gives the expression for the energy density $\epsilon$,
\begin{eqnarray}
\epsilon=\frac{E}{V}=\frac{1}{\pi^2\hbar^3}\int_{0}^{p_F}\frac{p^2\epsilon_pdp}{\left[1-\beta p+\gamma p^2\right]^4},
\end{eqnarray}
with $\epsilon_p=\sqrt{p^2c^2+m^2c^4}$ is the dispersion relation. For the pressure, by using the thermodynamics relation
\begin{eqnarray}\label{termo}
P=n\frac{\partial\epsilon}{\partial n}-\epsilon,
\end{eqnarray}
we obtain
\begin{eqnarray}\label{Pressure1}
P\left(p_F\right)=\frac{1}{\pi^2\hbar^3}\int_{0}^{p_F}\frac{p^2dp}{\left[1-\beta p+\gamma p^2\right]^4}\left[\epsilon_{F}-\epsilon_p\right],\nonumber \\
\end{eqnarray}
with $\epsilon_F=\sqrt{p_{F}^{2}c^2+m^2c^4}$ is the Fermi energy.

For the ultra-relativistic regime, where the kinetic energy is much higher than the rest energy $pc\gg mc^2$, the dispersion relation then can be approximate by $\epsilon_p\approx pc$, which gives the expression for the ultra-relativistic limit of the pressure or pressure at high momentum as
\begin{eqnarray}\label{ul}
P_{\text{ul}}\left(p_{F}\right)&=&\frac{c}{\pi^2\hbar^3}\int_{0}^{p_F}\frac{p^2dp}{\left[1-\beta p+\gamma p^2\right]^4}\left(p_F-p\right)\nonumber \\
&=&-\frac{c}{3\pi^2\hbar^3\left(4\gamma-\beta^2\right)^3}\left[G\left(p_{F},\beta,\gamma\right)+H\left(p_{F},\beta,\gamma\right)\right],
\end{eqnarray}
with
\begin{eqnarray}
G\left(p_{F},\beta,\gamma\right)&=&11\beta^2+16\gamma-\beta^3p_{F}-26\beta\gamma p_{F}\nonumber\\
&+& \frac{6\left(\beta^3+6\beta\gamma-4\beta^2\gamma p_{F}-4\gamma^2 p_{F}\right)\tan^{-1}\left(\frac{\beta}{\sqrt{4\gamma-\beta^2}}\right)}{\sqrt{4\gamma-\beta^2}},\nonumber \\
\end{eqnarray}
and
\begin{eqnarray}
H\left(p_{F},\beta,\gamma\right)&=&-\frac{\left(\beta^2-4\gamma\right)^2\left(\beta^2-2\gamma\right)}{\gamma^2\left(1-\beta p_{F}+\gamma p_{F}^{2}\right)^2}-\frac{3\left(\beta-2\gamma p_{F}\right)\left(\beta^3+6\beta \gamma-4\beta^2\gamma p_{F}-4\gamma^2 p_{F}\right)}{2\gamma\left(1-\beta p_{F}+\gamma p_{F}^{2}\right)} \nonumber \\
&&-\frac{\left(4\gamma-\beta^2\right)\left(2\beta^4-3\beta^3\gamma p_{F}-8\beta\gamma^2 p_{F}+4\gamma^2\left(6+\gamma p_{F}^{2}\right)+\beta^2\gamma \left(-9+4\gamma p_{F}^{2}\right)\right)}{2\gamma^2\left(1-\beta p_{F}+\gamma p_{F}^{2}\right)}\nonumber \\
&&+\frac{6\left(\beta^3+6\beta\gamma-4\beta^2\gamma p_{F}-4\gamma^2 p_{F}\right)\tan^{-1}\left(\frac{-\beta+2\gamma p_{F}}{\sqrt{4\gamma-\beta^2}}\right)}{\sqrt{4\gamma-\beta^2}},\nonumber \\
\end{eqnarray}

Note that $P_{\text{ul}}$ is also a transcendental function of the Fermi momentum.  Based on this fact, we cannot also express the EOS relating pressure to number density in a polytropic form. An extreme condition can be consider by taking $p_F$ approach infinity ($p_F\rightarrow\infty$). In this limit, the result yields a constant number density in Eq. (\ref{den})
\begin{eqnarray}\label{den1}
n\left(p_{F}\rightarrow\infty\right)=\frac{1}{3\pi^2\hbar^3\left(4\gamma-\beta\right)^3}\left[\beta^3+26\beta\gamma+\frac{12\gamma\left(\beta^2+\gamma\right)}{\sqrt{4\gamma-\beta^2}}\left(\pi+2\tan^{-1}\left(\frac{\beta}{\sqrt{4\gamma-\beta^2}}\right)\right)\right],\nonumber \\
\end{eqnarray}
and in this limit we can related with a standard linear EOS as follow
\begin{eqnarray}
P_{\text{ul}}\left(p_F\rightarrow\infty\right)&=&n\left(p_{F}\rightarrow\infty\right)c p_{F}+\frac{2c\left(\beta^2+\gamma\right)}{3\pi^2\hbar^3\left(4\gamma-\beta^2\right)^3\gamma}\left[6\gamma+\left(4\gamma-\beta^2\right)\right].
\end{eqnarray}
It means in the limit of $p_F\rightarrow\infty$, the pressure is softer compared to that of standard Fermi gas. These results demonstrate that this kind of GUP model provide EOS with similar behaviour with the one obtained for the quadratic GUP model. Therefore, we can conclude that the existence of minimal length in general softens the pressure at the ultra-relativistic regime. For  linear GUP model, one can directly write the number of states expression as \cite{ali2011minimal}

\begin{equation}\label{lqg2}
\Omega=\frac{g}{\left(2\pi\right)^3\hbar^3}\int \frac{d^3xd^3p}{\left(1-\beta p\right)^4}.
\end{equation}
We obtain the expression for the number density $n=\frac{N}{V}$ as
\begin{eqnarray}\label{den2}
n\left(p_F\right)&=&\frac{1}{\pi^2\hbar^3}\int_{0}^{p_{F}}\frac{p^2dp}{\left(1-\beta p\right)^4}\nonumber \\
&=&\frac{1}{3\pi^2\hbar^3}\frac{p_{F}^{3}}{\left(1-\beta p_F\right)^3}.
\end{eqnarray}
Since $n\left(p_F\right)$ is not a transcendental function, we can express the Fermi momentum as a function of the number density as
\begin{eqnarray}\label{den3}
p_{F}&=&\frac{C_1}{1+\beta C_1n^{\frac{1}{3}}}n^{\frac{1}{3}}\approx C_1\left(1-\beta C_1n^{\frac{1}{3}}\right)n^{\frac{1}{3}},
\end{eqnarray}
where $C_1=\left(3\pi^2\right)^{\frac{1}{3}}$. Based on this relation, we can express the EOS of white dwarfs  at the extreme regime as a polytropic. By writing the density energy as
\begin{eqnarray}
\epsilon=\frac{1}{\pi^2\hbar^3}\int_{0}^{p_F}\frac{p^2\epsilon_pdp}{\left[1-\beta p\right]^4},
\end{eqnarray}
and using the thermodynamics relation in Eq.(\ref{termo}) we obtain the expression for the pressure as
\begin{eqnarray}\label{Pressure}
P\left(p_F\right)=\frac{1}{\pi^2\hbar^3}\int_{0}^{p_F}\frac{p^2dp}{\left[1-\beta p\right]^4}\left[\epsilon_{F}-\epsilon_p\right].
\end{eqnarray}
For the ultra-relativistic regime, where $\epsilon_p\approx pc$, we obtain the ultra-relativistic limit of the pressure as a function of the Fermi momentum to first order in $\beta$ as
\begin{eqnarray}
P_{\text{ul}}\left(p_{F}\right)=\frac{c}{\pi^2\hbar^3}\left[\frac{p_{F}^{4}}{12}+\frac{\beta}{5}p_{F}^{5}\right].
\end{eqnarray}
By substituting Eq. (\ref{den3}) on the above equation, we have
\begin{eqnarray}
P_{\text{ul}}\left(n\right)=\frac{c}{\pi^2\hbar^3}\left[\frac{C_{1}^{4}}{12}n^{\frac{4}{3}}-\frac{2\beta}{15}C_{1}^{5}n^{\frac{5}{3}}\right].
\end{eqnarray}
Since for the white dwarfs, energy density is dominated from the electrons, then $\epsilon=\nu_em_{N}n$, where $\nu_e$ are the number of nucleons per electron and $m_N$ are the nucleon mass. Based on this we can express the pressure as a function of the energy density as
\begin{eqnarray}
P_{\text{ul}}\left(\epsilon\right)=\frac{c}{\pi^2\hbar^3}\left[\frac{C_{1}^{4}}{12\left(\nu_em_N\right)^{\frac{4}{3}}}\epsilon^{\frac{4}{3}}-\frac{2\beta}{15\left(\nu_em_N\right)^{\frac{4}{3}}}C_{1}^{5}\epsilon^{\frac{5}{3}}\right].\nonumber \\
\end{eqnarray}
The first correction above is the usual polytropic term, with polytropic index $n=3$. The modified phase space volume inclusion introduces a correction term, which we take only to the first order. The correction term is also behaving as a polytrope but with a different polytropic index, i.e.,  $n=\frac{3}{2}$.

To this end, we can conclude that the GUP corrections in EOS as the second ingredient needed for calculating mass-radius relations provide richer analytic structures than that of the standard one at enormous Fermi momentum value. Therefore, we are interested in checking if these structures might leave remnants at low densities or be imprinted on the properties of white dwarfs.
 
\subsection{White dwarfs mass and radius}
\label{s7}
For the non-relativistic stars, the hydrostatic equation with the proper EOS is sufficient to describe their mass-radius relation. White dwarfs are non-relativistic stars. Even though the degenerate electrons' behavior inside white dwarfs is relativistic, their pressure is far less than their corresponding energy density \cite{glendenning2012compact}. Therefore, it is unnecessary to use the Tolman-Oppenheimer-Volkoff equation to describes its pressure profiles. This section discusses the mass-radius relation for the white dwarfs by including the corrections from the GUP models. Here, we perform twofold modifications, i.e., the hydrostatic equation discussed in Sec. \ref{s3} and the EOS, which is discussed in Subsec. \ref{jiah}. We solve the hydrostatic equilibrium equation numerically using the Runge Kutta fourth-order algorithm and the EOS as input for arbitrary white dwarf pressure. We integrate the equations from the center pressure $p_c$ until the pressure vanishes on the star's surface. The products of integration of the corresponding equation are the radii and masses of white dwarfs. The dimension-free parameters of GUP models used before being transformed into the ones with the same dimension make numerical calculations easier. It means, the corresponding free parameters are calibrated into the Planck mass units such as $\beta=\frac{\beta_0}{M_{P}c}$ and $\gamma=\frac{\gamma_0}{M_{P}^{2}c^2}$. 
To see how compatible the GUP models with observation data, we parametrize the GUP free parameters with observational data using the $\chi^2$ analysis. The data are twelve observational mass-radius of white dwarf data collected in Ref. \cite{holberg2012observational}. We also show the observation data points and their deviations on each of our mass-radius figures for comparison. The fitting procedure is based on the following minimized definition.
\begin{equation}
\Delta\chi_i^2=\frac{\left(M-M_i\right)^2}{\sigma_{M,i}^2}+\frac{\left(R-R_i\right)^2}{\sigma_{R,i}^2},
\end{equation}
with $M_i$, $R_i$, $\sigma_{M_i}$, and $\sigma_{R_i}$ are the observational mass and radius data, as well as the corresponding standard deviation of mass and radius data, respectively. Whereas $M$ and $R$ are the mass and radius obtained from calculation. For each observation data, we can extract the minimum $\Delta\chi_i^2$ for a specific free parameter value. Calculate $\chi^2=\sum_{i}{\Delta\chi_i^2}$ and then run for a different parameter value to obtain the relations between the free parameter as a function of $\chi^2$. Therefore, we can extract the minimum $\chi^2$, for a certain free parameter value where the mass-radius relation calculated by using this corresponding free parameter is one of the most compatible with the observational data points. To identify our fitting's confidence level, we also plot the probabilities/likelihoods as functions of free parameters. The probability equation is defined as
\begin{equation}\label{prob}
P\left(\nu,\chi^2\right)=\int_{\chi^2}^{\infty}{\frac{2^{-\frac{\nu}{2}}}{\Gamma\left(\frac{\nu}{2}\right)}e^{-\frac{x}{2}}x^{\frac{\nu}{2}-1}dx}=\frac{\Gamma\left(\frac{\nu}{2},\frac{\chi^2}{2}\right)}{\Gamma\left(\frac{\nu}{2}\right)},
\end{equation}\nonumber \\
where $\nu$ is the number of degree of freedom, we define $\nu$ as $2 N_{o}-N_{f}-1$ with $N_o$ and $N_f$ are the number of observational data and fitting variable, while the factor of $2$ comes from two kinds of data, i.e., mass and radius. It is worthy to note that the probability scale ranges from zero to one. We have the best compatibility with the observational data when the probability is closer to one.

\begin{figure}
\includegraphics[width=70mm,scale=2.5]{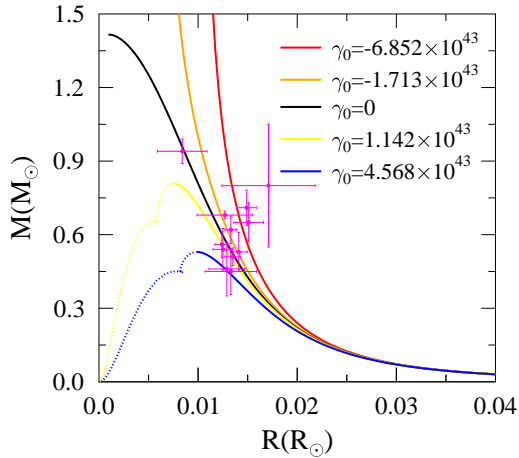}
\caption{\label{fig:tovq} The mass-radius relation of white dwarfs with quadratic GUP correction. We plot the mass and radius in solar units. Every point after the one with maximum mass but with a smaller radius than that of maximum mass in the positive value of $\gamma_0$ results represent the areas of unstable stars. }
\end{figure}

\begin{figure}
\includegraphics[width=70mm,scale=2.5]{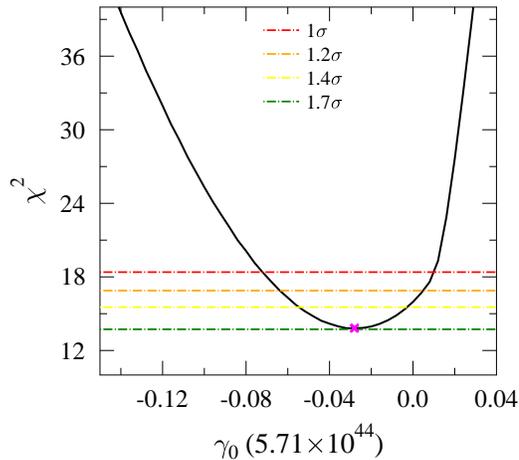}
\caption{\label{fig:chiq} The $\chi^2$ as a function of $\gamma_0$ for the quadratic GUP model. The dashed lines represent the confidence level of our fitting.}
\end{figure}

\subsection{Mass-radius in quadratic GUP model}
This subsection describes the impact of the $\gamma_0$ parameter on white dwarfs mass-radius relations predicted by the quadratic GUP model. Here we constrain the dimensionless $\gamma_0$ using the $\chi^2$ analysis. The mass-radius plots for some $\gamma_0$ values of the quadratic GUP model are shown in Fig. \ref{fig:tovq}. The increase in $\gamma_0$ causes a decrease in mass for a fixed radius value. It means the star's compactness consequently will also decrease. Unlike in a positive value of $\gamma_0$ case, for the negative value of $\gamma_0$ case, there is no indication of maximum mass or mass limit. However, for a positive value of $\gamma_0$, the mass limit is lower than the Chandrasekhar mass limit. It can also be seen that the mass and radius relation for a negative value of $\gamma_0$ is more closer to observational data. Note that, especially for the low-density regime (relative extensive R regime), we find the change in $\gamma_0$ value does not influence the mass-radius relation significantly.

\begin{figure}
\includegraphics[width=80mm,scale=2.5]{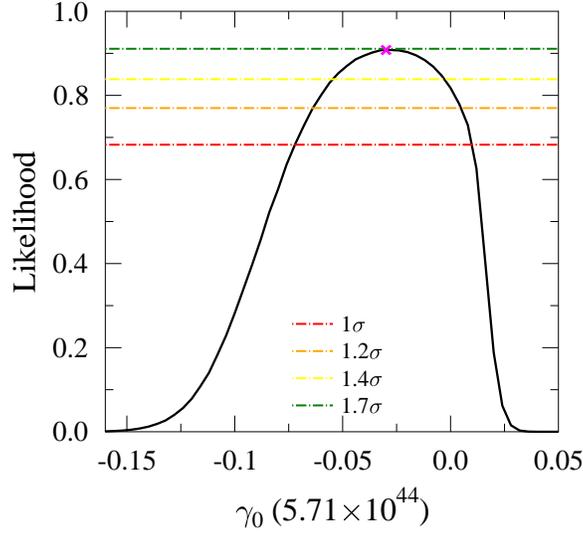}
\caption{\label{fig:probq}The probability as a function of $\gamma_0$ for the quadratic GUP model. The dashed lines represent the level of confidence in our fittings.}
\end{figure}

Figs. \ref{fig:chiq} and \ref{fig:probq} show the $\chi^2$ and the probability as a function of $\gamma_0$ predicted by quadratic GUP model, respectively. We obtain that the minimum value of $\chi^2$ is $13.81$, and the maximum probability value is $0.91$, which occurs when $\gamma_0=-1.656\times10^{43}$. There are no other extremum points that appeared in the figure. Therefore, it is clear that $\gamma_0=-1.656\times10^{43}$ is the best $\gamma_0$ value, which is the corresponding mass and radius predictions compatible with observational data. We also obtain that the corresponding optimal $\gamma_0$ has a confidence level of $1.7\sigma$. Furthermore, this best fitted to the data of $\gamma_0$ value predicts no mass limit. This result is quite contrasted if we compare it to that of white dwarfs within the standard description.

\begin{figure}
\includegraphics[width=80mm,scale=2.5]{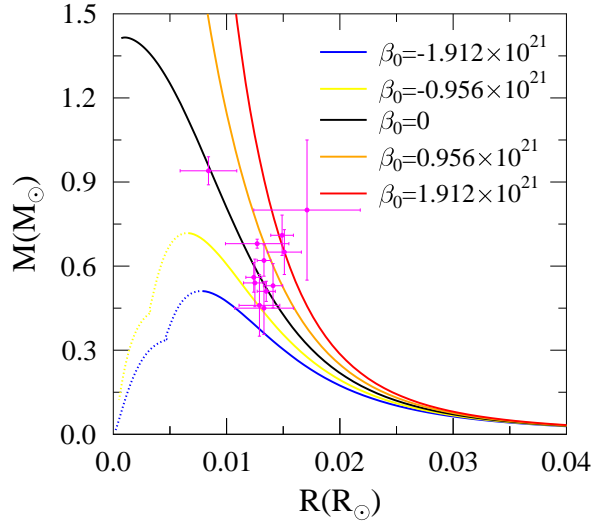}
\caption{\label{fig:tovl} The mass-radius relation of white dwarfs with linear GUP correction. We plot the mass and radius in solar units. Every point after the one with maximum mass but with a smaller radius than that of maximum mass in a negative value of $\beta_0$ represent the areas of unstable stars. }
\end{figure}

\begin{figure}
\includegraphics[width=80mm,scale=2.5]{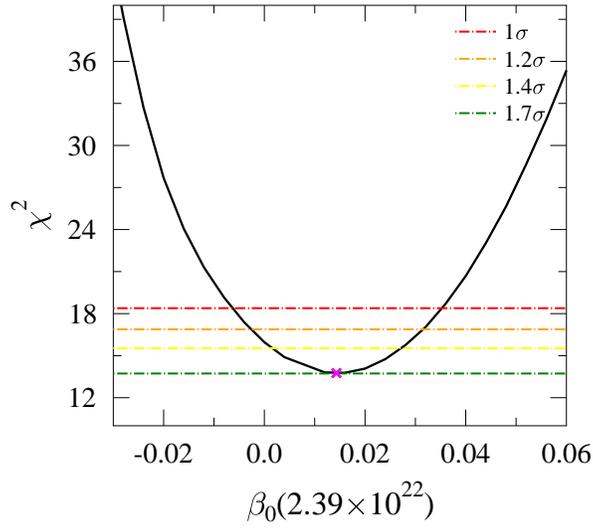}
\caption{\label{fig:chil}The $\chi^2$ as the function of $\beta_0$ for the linear GUP model. The dashed lines represent the level of confidence in our fittings.}
\end{figure}

\begin{figure}
\includegraphics[width=80mm,scale=2.5]{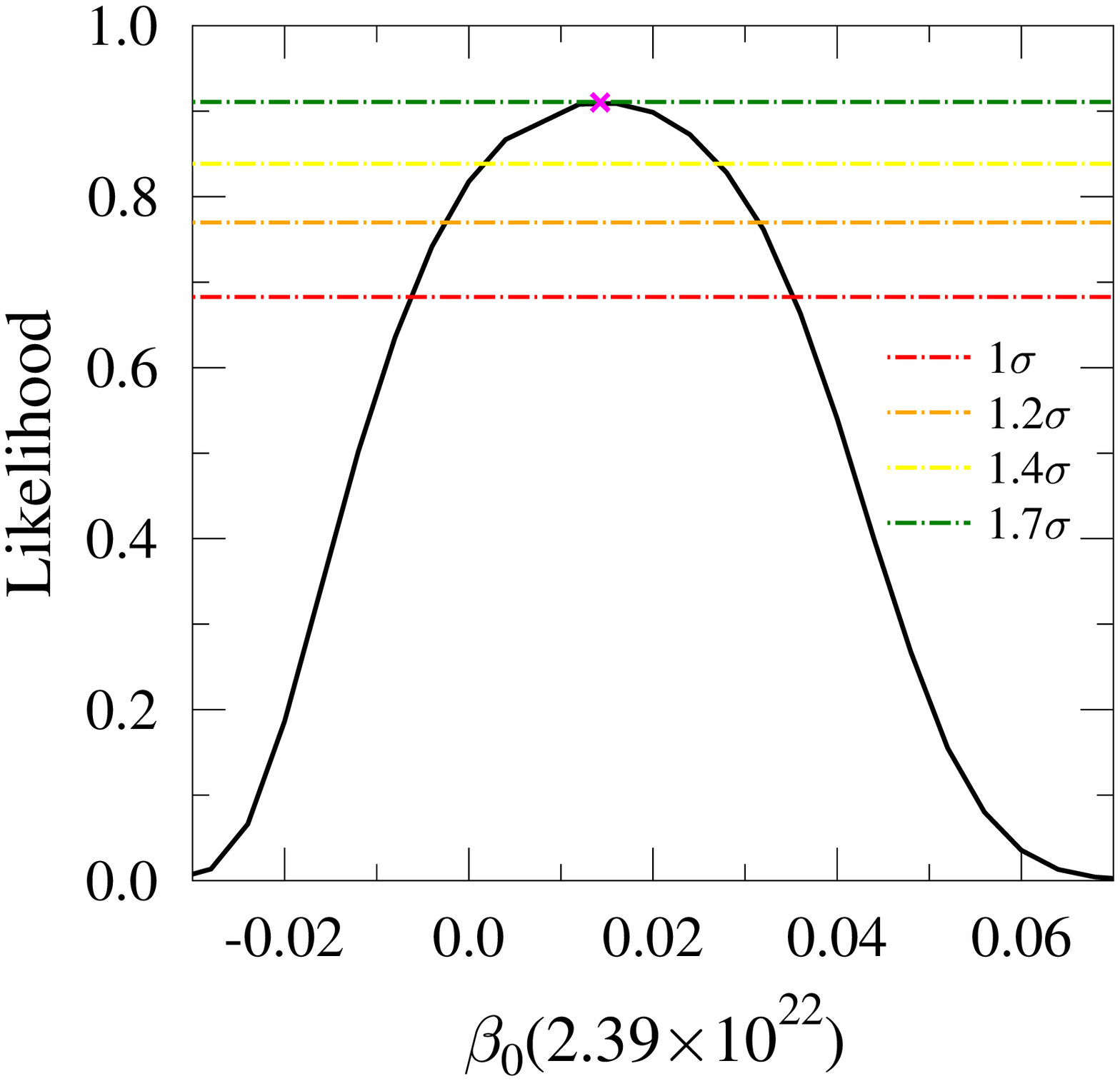}
\caption{\label{fig:probl} The probability values as a function of $\beta_0$ for the linear GUP model. The dashed lines represent the level of confidence in our fittings.}
\end{figure}

\subsection{Mass-radius in linear GUP Model}
The white dwarfs mass-radius relation predicted by the linear GUP model is shown in Fig. \ref{fig:tovl}. The increasing $\beta_0$ causes the mass increase for a fixed radius value. It means that the star's compactness will increase too if $\beta_0$ is increased. There is no maximum mass limit found for a positive $\beta_0$ value. However, for a nonzero but negative $\beta_0$ value, the mass limit is smaller than that of Chandrasekhar. It can be concluded that the trend of mass-radius relation predictions of the linear model is quite similar to that of the quadratic GUP model, including the behavior of mass-radius relation at the low-density regime. However, the changes due to the sign of $\beta_0$ value in the linear GUP model on mass-radius relations has the opposite effect compared to the sign changes in $\gamma_0$ of the quadratic GUP model.

Figs. \ref{fig:chil} and \ref{fig:probl} display the $\chi^2$ and the probability as a function of $\beta_0$ predicted by linear GUP model, respectively. We obtain the minimum $\chi^2$ is $13.75$, and the maximum probability is $0.91$, which occurs when $\beta_0=3.346\times10^{20}$. The highest confidence level is still almost the same quality as the one of the quadratic GUP model, where the obtained value of around $1.7\sigma$. Furthermore. Like that of the quadratic GUP model, the best fitting parameter also does not provide the mass limit.

\begin{figure}
\includegraphics[width=90mm,scale=2.6]{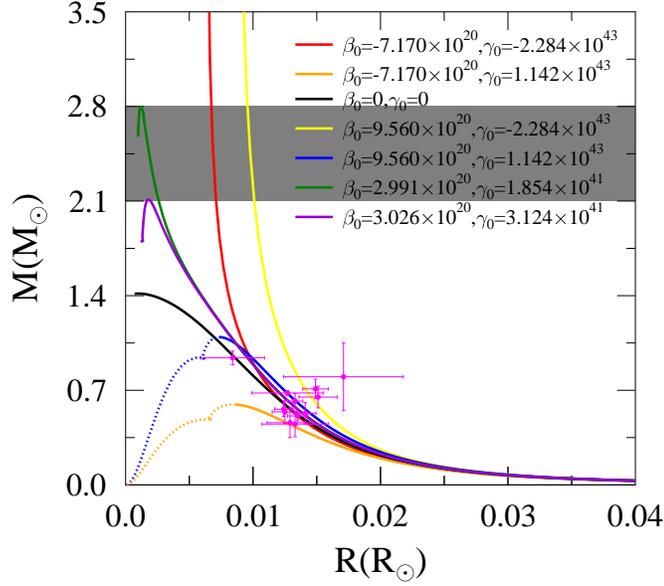}
\caption{\label{fig:tovlq}The mass-radius relation of a white dwarf with a linear-quadratic GUP model. We plot the mass and radius in solar units. Every point after the one with maximum mass but with a smaller radius than that of maximum mass represents the areas of unstable stars. Grey shaded area is a constraint from the ultra massive white dwarfs observation data.}
\end{figure}

\subsection{Mass-radius in linear-quadratic GUP Model}\label{KKl}

\begin{figure}
\includegraphics[width=80mm,scale=2.5]{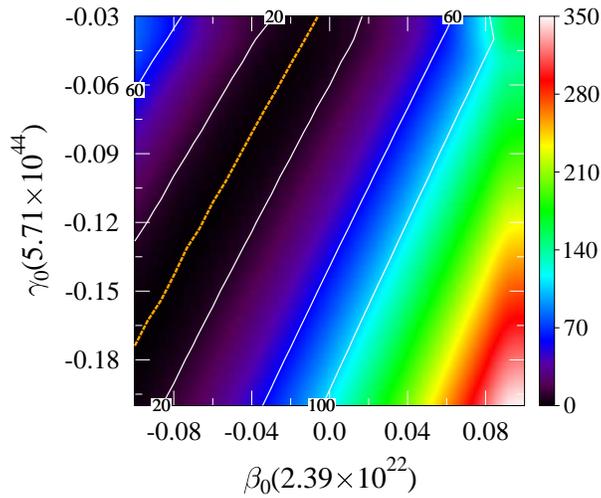}
\caption{\label{fig:chilq}The contour plot of $\chi^2$  as a function of $\beta_0$ and $\gamma_0$.}
\end{figure}

\begin{figure}
\includegraphics[width=80mm,scale=2.5]{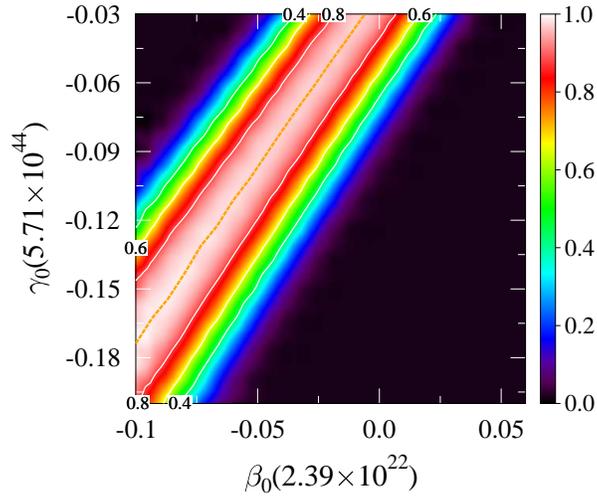}
\caption{\label{fig:problq} The contour plot of probability as a function of $\beta_0$ and $\gamma_0$ .}
\end{figure}

The plot of the mass-radius relations for the phenomenological extension form of the linear-quadratic GUP model is shown in Fig. \ref{fig:tovlq}. The change of mass in this GUP model can be controlled by inter-playing between $\beta_0$ and the role of $\gamma_0$ parameters. $\beta_0<0$ or $\gamma_0>0$ makes the mass decrease, and $\beta_0>0$ or $\gamma_0<0$ makes the mass increase. Therefore, in the linear-quadratic GUP model, we still have the flexibility to have a larger mass limit than that of the standard Chandrasekhar mass limit, but the corresponding prediction still compatible with the observational data of masses and radii of some standard white dwarfs by adjusting $\beta_0$ and $\gamma_0$ parameters, simultaneously. This feature is reflected in Fig.\ref{fig:tovlq}. It can be seen that the ultra massive white dwarfs which are recently observed can be nicely predicted within this type of linear-quadratic GUP model. This GUP model also predicts a relatively small radius for the maximum mass larger than the standard Chandrasekhar mass limit. Meanwhile, at the low-density regime, the mass does not depend significantly on both free parameter values. Similar to that of the quadratic GUP model, It can observe in Fig.\ref{fig:tovlq} that in the case of the negative value of $\gamma_0$ parameter, the mass limit is not present. However, by increasing the positive $\beta_0$ and decreasing the positive $\gamma_0$ parameter, we can obtain a limiting mass larger than the Chandrasekhar mass limit.  However, if the values of $\beta_0$ and  $\gamma_0$ parameters relative too large, we obtain the mass limit smaller than the Chandrasekar mass limit.  This interplay is why we can not yield a good prediction of white dwarfs with massive white dwarfs mass data included with  $\beta_0^2$= $\gamma_0$. It is interesting to observe that this model's optimal prediction has an absolute $\gamma_0$ value two orders of magnitude smaller than that of the quadratic GUP model, and most importantly, the sign of the optimal $\gamma_0$ value is positive. However, If we compare the optimal $\gamma_0$ value predicted by this GUP model is still too large than those of best bounds of  $\gamma_0$ discussed in Ref.~\cite{Scardigli2019}. However, we need to note that our best $\beta_0$  value is in the same order, but the best $\gamma_0$ value is smaller than that obtained from constraining the GUP parameter using gravitational wave event GW150914~\cite{Feng2017,Scardigli2019}. It means that if the gap of the free GUP parameter bounds~\cite{Scardigli2019} is narrower in the future with the order of magnitude smaller than that obtained in this work, the existence of massive white dwarfs becomes unlikely explained by GUP corrections. 

Figs. \ref{fig:chilq} and \ref{fig:problq} show the impact of the variation of $\beta_0$ and $\gamma_0$ on $\chi^2$ and likelihood probability, respectively. We present the $\chi^2$ and probability as a function of $\beta_0$ and $\gamma_0$ parameters in contour plot forms. It can be seen that a linear relation between $\beta$ and $\gamma_0$ with the same quality of $\chi^2$ or probability exists. It means there are some sets of $\beta_0$ and $\gamma_0$ parameter combinations with the same quality of $\chi^2$ or probability. The points in line with minimum $\chi^2$ or maximum probability can be obtained, and they are plotted as an orange dashed line in both figures. We can parametrize all points in this line as a linear equation. We obtain that the line obeys the following relation:
\begin{equation}\label{rt0}
  \frac{\gamma_0}{5.71\times10^{44}}=-0.019+1.544\frac{\beta_0}{2.39\times10^{22}}.
\end{equation}
  This line's confidence level is around $1.6\sigma$, with an average of $\chi^2$ and a probability of about $13.75$ and $0.88$, respectively. With this, we can search two points in the corresponding line that predict the upper and lower bound of maximum mass or mass limit larger than that of Chandrasekhar. This result can be seen in Fig.\ref{fig:tovlq}. Note that for standard linear-quadratic GUP model with  $\beta_0^2$= $\gamma_0$, the solution are $\beta_0= 2. 94\times 10^{19}$ and $\beta_0= 3. 66 \times10^{19}$, respectively. The first one has a mass limit relatively larger than Chandrasekar mass limit but still far from the mass of massive white dwarfs, due to too small the corresponding value of $\gamma_0$, while the second one yields a smaller mass limit compared to Chandrasekar mass limit. To this end, it is evident that the mass-radius data of white dwarfs prefer the phenomenological linear-quadratic GUP model with $\beta_0^2$ $\neq$ $\gamma_0$. However, the commutation relation in the phenomenological linear-quadratic GUP model used in this work is phenomenologically defined. Therefore, It needs further work to obtain the correct fundamental commutation relation related to the linear-quadratic GUP model because the corresponding Heisenberg algebra is not simple, and it should be carefully taken care of (see the Appendix of Ref. ~\cite{ali2011proposal} for detail about this matter). We leave this issue for our next work.

\section{Conclusion}
\label{s8}
GUP is a model to incorporate the existence of minimal length in quantum mechanics (i.e. the quantum gravity effect on the quantum mechanical system). The EOS and the gravitational field expressions are deformed due to the existence of the minimal length. Here, we derived the deform EOS of degenerate Fermi gas using the modified invariant phase-space volume and quantum corrected hydrostatic equation using Verlinde's method and investigate the impacts of these modifications on the mass-radius relation of the white dwarfs. To examine GUP models' compatibility with masses and radii of white dwarf observation data, we parametrize the free parameter of the corresponding model with the recent masses and radii of white dwarfs. In general, we have found that the GUP models are relative more compatible with data compared to the one of standard description, and the quality of the highest confidence levels of quadratic, linear, and linear-quadratic GUP models is similar. However, for the quadratic GUP model, data favors the negative value of $\gamma_0$, while For linear GUP model favors the positive value of $\beta_0$.
Consequently, masses and radii of white dwarfs data are more compatible with the white dwarf description without mass limit. This result confirms previous results \cite{rashidi2016generalized,ong2018generalized,mathew2018effect} that mass-radius relation for the white dwarfs within linear-quadratic GUP models is behaving differently from the standard one. For the quadratic-linear GUP model, we can obtain a positive value of  $\gamma_0$ parameter, and the model yields a mass limit that is larger than the Chandrasekar mass limit. However, our result indicates that we need   $\beta_0^2$ $\neq$ $\gamma_0$ to explained the massive white dwarfs.

\section*{Acknowledgments}
AS is partly supported by DRPM UI's (PUTI-Q1 and PUTI-Q2) grants No:NKB-1368/UN2.RST/ HKP.05.00/2020 and No:NKB-1647/UN2.RST /HKP.05.00/ 2020.

\end{document}